\documentclass[conference,a4paper]{IEEEtran}

\usepackage[utf8]{inputenc} 
\usepackage[T1]{fontenc}
\usepackage{url}
\usepackage{ifthen}
\usepackage{cite}
\usepackage[cmex10]{amsmath} 
\usepackage{amssymb}
\usepackage[linesnumbered,ruled]{algorithm2e}
\usepackage{graphicx}
\usepackage{multirow}

\newtheorem{thm}{Theorem}[section]
\newtheorem{lem}[thm]{Lemma}

\newtheorem{defn}[thm]{Definition}

\newtheorem{exmp}[thm]{Example}

\newtheorem{claim}{Claim}

\begin{document}
\title{On the Structure of Interlinked Cycle Structures with Interlocked Outer Cycles}
\author{\IEEEauthorblockN{Shanuja Sasi and B. Sundar Rajan} 
\IEEEauthorblockA{Dept. of Electrical Communication Engg., Indian Institute of Science Bangalore, Karnataka, India - 560012\\
			Email: \{shanuja,bsrajan\}@iisc.ac.in}}	
\maketitle
 \thispagestyle{plain}
\pagestyle{plain}
	
\begin{abstract}

For index coding problems with special structure on the side-information graphs called Interlinked Cycle (IC) structures index codes have been proposed in the literature (C. Thapa, L. Ong, and S. Johnson, ``Interlinked Cycles for Index Coding: Generalizing Cycles and Cliques", in \textit{IEEE Trans. Inf. Theory, vol. 63, no. 6, Jun. 2017} with a correction in ``Interlinked Cycles for Index Coding: Generalizing Cycles and Cliques", in arxiv (arxiv:1603.00092v2 [cs.IT] 25 Feb 2018)). Recently (S. Sasi and B.S. Rajan, ``On Optimal Index Codes for Interlinked Cycle Structures with Outer Cycles,'' in arxiv (arXiv:1804.09120v1 [cs.IT]), 24 Apr 2018) for a generalization of IC structures called {\it IC structures with interlocked outer cycles} optimal length index codes have been reported and it is shown that the optimal length depends  on the maximum number of disjoint outer cycles. In this paper we discuss certain structural properties of IC structures with interlocked outer cycles and provide a simple  algorithm to find the maximum number of disjoint outer cycles. 
\end{abstract}
\section{INTRODUCTION}
\subsection{Background and Preliminaries}
In a single sender Index Coding problem with Side Information (ICSI), there will be a unique source having a set of $P$ messages $\mathcal{M} = \{x_{1},x_{2},...,x_{P}\}$ and $N$ receivers $\mathcal{R} =\{\mathcal{R}_{1},\mathcal{R}_{2},...,\mathcal{R}_{N} \}$. Each receiver $\mathcal{R}_i$ is characterized by $(\mathcal{W}_i,\mathcal{K}_i)$, where $\mathcal{W}_i \subseteq \mathcal{M}$ is the set of messages it demands and $\mathcal{K}_i \subseteq \mathcal{M}$ is the set of messages it knows a priori which is also the side information possessed by it, where $\mathcal{W}_i\cap\mathcal{K}_i=\emptyset$, i.e, it doesn't demand anything which it already has. The sender knows the side information available to the receivers. Each receiver should get the messages it requested from the sender with minimum number of transmissions. An ICSI is identified by $(\mathcal{M},\mathcal{R})$. An index code for the ICSI ($\mathcal{M},\mathcal{R}$) consists of an encoding function for the sender, $\mathcal{E} : \mathbb{F}_{q}^P \rightarrow \mathbb{F}_{q}^l$ and a set of decoding functions $\mathcal{D}_i : \mathbb{F}_{q}^l \times \mathbb{F}_{q}^{|\mathcal{K}_i|} \rightarrow \mathbb{F}_{q}^{|\mathcal{W}_i|}$ for each $i \in \{1,2,..,N\} $, such that,
			\begin{equation}
				\mathcal{D}_i(\mathcal{E}(\mathcal{\mathcal{M}}),\mathcal{K}_i) = \mathcal{W}_i ,\forall i \in \{1,2,..,N\},
			\end{equation}
			where $l$ is the length of the index code and $\mathbb{F}_q$ is a finite field with $q$ elements. The objective is to find the optimal index code which has the smallest $l$ possible, such that each receiver can decode the messages, which it demanded, from the codeword transmitted and the known messages. An index code is said to be linear if the encoding function $\mathcal{E}$ is linear. ICSI problem was introduced in \cite{BiK}.
		
			A detailed classification of ICSI is done by Ong and Ho \cite{OngHo} based on the demands and side information of the receivers. An ICSI problem is said to be \textit{unicast} if $\mathcal{W}_i \cap \mathcal{W}_j = \emptyset$ $\forall$ $i \neq j $ and it is \textit{single unicast} if $|\mathcal{W}_i|=1$. An ICSI problem is said to be \textit{uniprior} if $\mathcal{K}_i \cap \mathcal{K}_j = \emptyset$ $\forall$  $i \neq j $ and it is \textit{single uniprior} if $|\mathcal{K}_i|=1$. The most general setting of index coding instance is when there is no restriction on $\mathcal{W}_i$ and $\mathcal{K}_i$, which is said to be \textit{multicast multiprior}. Finding the optimal length for this case is known to be NP hard \cite{AHL}. 
			
			The notation $\lfloor v \rceil$ is used to represent the set $\{1,2,...,v\}$ for any integer $v$. We consider single unicast problem with $P=N$. Without loss of generality, let each receiver $R_i$ requests the message $x_i$. A \textit{Digraph} $G$ is specified by a set of vertices $V(G)$ and a set of arcs $E(G)$ connecting the vertices, denoted by $(V, E)$. An arc $(i, j) \in E(G)$, if the receiver $R_i$ has $x_j$ as side information. A \textit{path} is a digraph $P_{v_a,v_b}$, specified by $(V, E)$, where $V= \{v_a,v_1,v_2,...,v_t,v_b\}$ and $E=\{(v_a \rightarrow v_1),(v_1\rightarrow v_2),(v_2\rightarrow v_3),...,(v_t\rightarrow v_b)\}$. It is represented by $P_{v_a,v_b} = (v_a \rightarrow v_1 \rightarrow v_2 \rightarrow...\rightarrow v_t \rightarrow v_b)$. Two paths from $v_a$ to $v_b$ are said to be distinct, if the two paths do not have any vertices in common other than $v_a$ and $v_b$. A single vertex is a path of length zero.
			
			The \textit{optimal broadcast rate} for an index coding problem $G$ is defined as $$\beta(G) = \inf_{t} {\beta_t(G)},$$ where $\beta_t(G)$ is the minimum number of index code symbols to be transmitted, in order to satisfy the demand of all the receiver, per t-bit messages. A \textit{Maximum Induced Acyclic Sub-Digraph (MAIS)} is the induced acyclic sub-digraph formed by removing minimum number of vertices from the digraph $G$. $MAIS(G)$ is the order of the $MAIS$. For any digraph $G$, $MAIS(G)$ is the lower bound on the optimal broadcast rate, i.e, $\beta(G) \geq MAIS(G)$.
			
			In \cite{TOJ1,TOJ2} the authors consider a digraph $G$ with $N$ vertices and define an Interlinked Cycle (IC) structure as follows. Let $V$ be the set of $N$ vertices. Let there be a vertex set $V_I$, such that it has $K$ vertices, $V_I=\{1,2,...,K\}$, where for all ordered pair of vertices ($i$, $j$), there exists exactly one path from $i$ to $j$ which doesn't include any other vertices in $V_I$ except $i$ and $j$, where $i,j \in V_I$ and $i \neq j$. Such a vertex set $V_I$ is called inner vertex set and the vertices in $V_I$ is called inner vertices. All other vertices in $G$ other than inner vertices are called non-inner vertices and are included in the vertex set $V_{NI}$. A path in which only the first and the last vertices are from $V_I$, and they are distinct, is called an I-path. If the first and the last vertices are same, then it is called an I-cycle.   
			
			There exists a directed rooted tree from every vertex in $V_I$, denoted by $T_i, $ with any inner vertex $i$ as the root vertex and all other vertices in $V_I$ other than $i$ as leaves. If the digraph $G$ satisfies the following four conditions, then it is called an Interlinked Cycle structure, denoted by $G_K$. 
\begin{itemize}
\item \textit{Condition 1:} The union of all the $K$ rooted trees, each one rooted at unique inner vertex, should form the digraph $G$.
\item \textit{Condition 2:} There is no I-cycle in $G_K$.
\item \textit{Condition 3:} For all ordered pair of vertices ($i$, $j$), there exists exactly one path from $i$ to $j$ which doesn't include any other vertices in $V_I$ except $i$ and $j$, where $i,j \in V_I$ and $i \neq j$.
\item \textit{Condition 4:} No cycle containing only non-inner vertices. \\
\end{itemize}   
	In \cite{TOJ1,TOJ2}, the authors have proposed an index code construction scheme for an IC structure $G_K$ with $K$ inner vertices and a total of $N$ vertices. The code construction is as follows:
\begin{itemize}
\item A coded symbol is obtained by bitwise XOR of messages requested by all the inner vertices in $V_I$.
\item A coded symbol is obtained by the bitwise XOR of the message requested by $j$ and the messages requested by all its out-neighborhood vertices in $G_K$, for all $j \in V_{NI}$.
\end{itemize}
The code constructed by the above procedure has $N-K+1$ coded symbols. \\

The cycles containing only the non-inner vertices are called outer cycles. The IC structures satisfying only the first three conditions are called IC structures with outer cycles. This paper deals with IC structures with outer cycles. In \cite{VaR}, optimal length index code is provided for IC structures with one cycle among non-inner vertex set. In \cite{ViR}, an index code of length $N-K+1$ is provided for IC structures with arbitrary number of outer cycles with the issue of optimality of the index codes left open. In our prior work\cite{SaR} we dealt with  optimal index code constructions for IC structures with any number of outer cycles satisfying two conditions calling such IC structures as {\it IC structures with interlocked outer cycles}. \\

\begin{defn}
\textit{Interlocked Outer Cycles\cite{SaR}:} Let there exist $n$ number of cycles, $C_j$ where $j \in \lfloor n \rceil$, among non-inner vertex set. The set of all cycles among non-inner vertices which obeys the two conditions below is called Interlocked Outer Cycles. Let $\cal{C}$ be the set of all outer cycles, $\cal{C}$ $=\{C_1,C_2,...,C_n\}$.
\begin{itemize}
\item \textit{Interlocking Condition (ILC):} If any cycle $C_i$ intersects any other cycle $C_k$, then $C_i$ and $C_k$ have exactly a path in common, for any $i,k \in \lfloor n \rceil$. 
\item \textit{Central Cycle Condition (CCC):} A cycle $C_c \in \cal{C}$ such that all other cycles in $\cal{C}$ have atleast a single vertex in common with $C_c$ and the intersection of any two cycles in $\cal{C}$, other than $C_c$, has atleast a single vertex in common with $C_c$ is called a Central Cycle. There exists atleast one central cycle in $\cal{C}$.
\end{itemize}
\end{defn}
		
Let \\
$V_{C_j}=$ subset of non-inner vertices of $C_j$, for  $j \in \lfloor n \rceil$ \\
$\hat{C}=\cup_{i=1}^{n} C_i, \\ 
V_{OC} = \cup_{i=1}^{n} V_{C_i}$, \\
$V_{i,j}=V_{C_i} \cap V_{C_j}$, for any $i,j \in \lfloor n \rceil$, \\
 $\tilde{V}_{C_j}=$  the subset of vertices in $V_{C_j}$ which doesn't intersect any other cycles in $\cal{C}$, i.e., $\tilde{V}_{C_j}= V_{C_j} \backslash (\cup_{i \neq j,i=1}^{n} V_{i,j}$), where $j \in \lfloor n \rceil$. \\ 

\begin{figure}[!t]
	\centering
	\includegraphics[width=15pc]{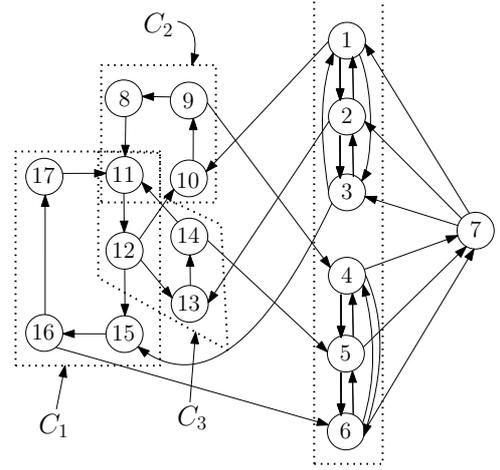}
	\caption{An IC structure $G_6$ with the path from the vertex $11$ to $12$ common to all the cycles in $C$.}
	\label{fig1}
\end{figure}

\begin{exmp}
	Consider the IC structure $G_{6}$ shown in Fig. \ref{fig1}. For this IC structure $N=17, K=6$, $V_I=\{1,2,3,4,5,6\}$, $V_{NI}=\{7,8,...,17\}$. Interlocked outer cycles are $C_1,C_2$ and $C_3$ with 
\begin{align*}
V_{C_1}&=\{11,12,15,16,17\}, \\
V_{C_2}&=\{11,12,10,9,8\}, \\
V_{C_3}&=\{11,12,13,14\},\\ 
 V_{1,2}&=\{11,12\}, \\ 
 V_{1,3}&=\{11,12\} \\ 
V_{2,3}&=\{11,12\}. 
\end{align*}

All the three cycles have the path from the vertex $11$ to $12$ in common. All the three cycles are central cycles.
	\label{exmp12}
\end{exmp}
\begin{figure}[!t]
	\centering
	\includegraphics[width=15pc]{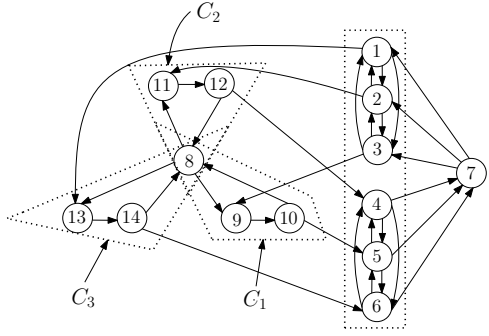}
	\caption{An IC structure $G_6$ with exactly a vertex $8$ common to all the cycles in $C$.}
	\label{fig2}
\end{figure}
\begin{exmp}
	Consider the IC structure $G_{6}$ shown in Fig. \ref{fig2}. For this IC structure $N=14, K=6$, $V_I=\{1,2,3,4,5,6\}$, $V_{NI}=\{7,8,...,14\}$. Interlocked outer cycles are $C_1,C_2$ and $C_3$ with 
\begin{align*}
V_{C_1}&=\{8,9,10\}, \\
V_{C_2}&=\{8,11,12\} \\
V_{C_3}&=\{8,13,14\}.
\end{align*}
 All of them have the vertex $8$ in common. All the three cycles are central cycles. 
\label{exmp13}
\end{exmp}
\begin{figure}[!t]
	\centering
	\includegraphics[width=20pc]{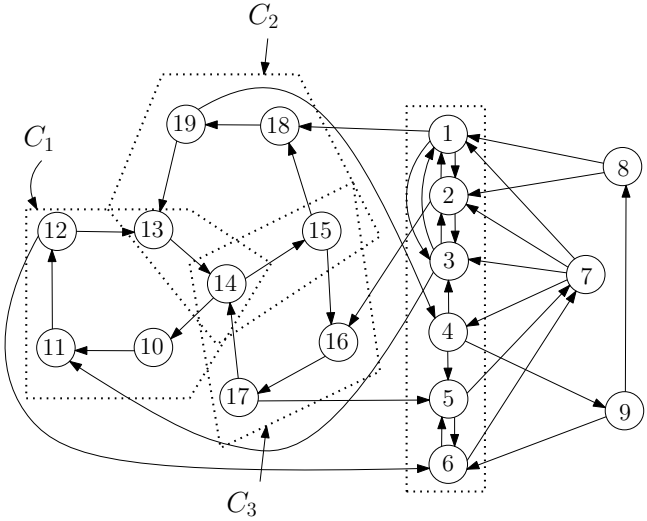}
	\caption{An IC structure $G_6$ with the vertex $14$ in common with all the cycles in $C$ and the vertex $13$ in common with $C_1$ and $C_2$ only.}
	\label{fig3}
\end{figure}


\begin{exmp}
	Consider the IC structure $G_{6}$ shown in Fig. \ref{fig3}. For this IC structure $N=19, K=6$, $V_I=\{1,2,3,4,5,6\}$, $V_{NI}=\{7,8,...,19\}$. Interlocked outer cycles are $C_1,C_2$ and $C_3$ with 
\begin{align*}
V_{C_1}&=\{10,11,12,13,14\},\\
V_{C_2}&=\{13,14,15,18,19\},\\
V_{C_3}&=\{14,15,16,17\}, \\
V_{1,2}&=\{13,14\}, \\ 
V_{1,3}&=\{14\}, \\
V_{2,3}&=\{14,15\}. 
\end{align*}
All the three cycles have the vertex $14$ in common. All the three cycles are central cycles.
	\label{exmp14}
\end{exmp}
\begin{figure}[!t]
	\centering
	\includegraphics[width=17pc]{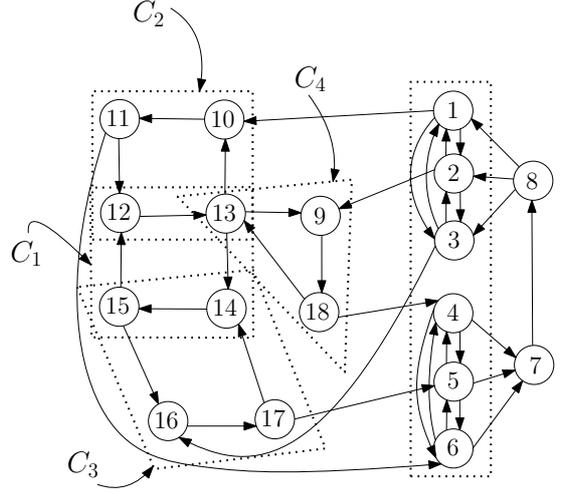}
	\caption{An IC structure $G_6$ which doesn't have any vertex in common with all the cycles in $C$.}
	\label{fig4}
\end{figure}

\begin{exmp}
	Consider the IC structure $G_{6}$ shown in Fig. \ref{fig4}. For this IC structure $N=18, K=6$, $V_I=\{1,2,3,4,5,6\}$, $V_{NI}=\{7,8,...,18\}$. Interlocked outer cycles are $C_1,C_2,C_3$ and $C_4$ with 
\begin{align*}
V_{C_1}&=\{12,13,14,15\},\\
V_{C_2}&=\{10,11,12,13\},\\
V_{C_3}&=\{14,15,16,17\},\\
V_{C_4}&=\{13,9,18\}, \\
V_{1,2}&=\{13,12\},\\
V_{1,3}&=\{14,15\},\\
V_{1,4}&=\{13\},\\
V_{2,3}&=\{13\}\\
V_{2,4}&=V_{3,4}= \phi.
\end{align*}
 All the four cycles do not have any vertex in common. The only central cycle is $C_1$.
	\label{exmp15}
\end{exmp}

\begin{figure}[!t]
	\centering
	\includegraphics[width=18pc]{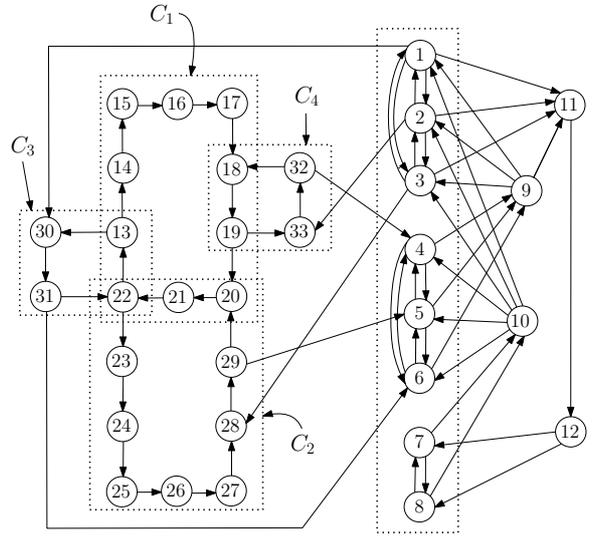}
	\caption{An IC structure $G_{8}$ which doesn't have any vertex in common with all the cycles in $C$.}
	\label{Fig5}
\end{figure}
\begin{exmp}
	Consider an IC structure $G_{8}$ shown in Fig. \ref{Fig5}. For this IC structure $N=33, K=8, V_I=\{1,2,3,4,5,6,7,8\}$ and $ V_{NI}=\{9,10,...,33\}$. Interlocked outer cycles are $C_1,C_2,C_3$ and $C_4$. 
\begin{align*}
V_{C_1}&=\{13,14,15,...,21,22\},\\
V_{C_2}&=\{20,21,22,...,28,29\}, \\
V_{C_3}&=\{13,30,31,22\},\\
V_{C_4}&=\{18,19,33,32\} ,\\
V_{1,2}&=\{20,21,22\}, \\
V_{1,3}&=\{22,13\},\\
V_{1,4}&=\{18,19\},\\
V_{2,3}&=\{22\}\\
V_{2,4}&=V_{3,4}= \phi.
\end{align*}
 There is no vertex in common with all the cycles. The only central cycle is $C_1$. \label{exmp16}\\	
\end{exmp}	

In \cite{SaR}, we established two properties of IC structures with interlocked outer cycles, i.e,
\begin{itemize}
	\item \textit{P1:} If all the $n$ cycles in $\cal{C}$ have atleast one vertex in common, then all the cycles in $\cal{C}$ are central cycles.

	\item \textit{P2:} For any central cycle $C_c \in \cal{C}$, there exists a path from any $i \in V_{C_k}$ to any $j \in V_{C_l}$, where $k,l \in \lfloor n \rceil$, and the path includes the vertices and arcs only from $C_k\cup C_l\cup C_c$. \\
\end{itemize}

 For IC structures with interlocked outer cycles, in \cite{SaR}, we have shown that the optimal length depends on the maximum number of disjoint outer cycles. To be precise, if $t$ denotes the maximum number of disjoint outer cycles then the optimal length is $N-K+2-t.$ We provided explicit optimal index code construction for the cases where either of the two conditions- Condition $\textit{C1}$ or $\textit{C2}$ - described below is satisfied. 

\begin{itemize}
        \item \textit{C1:} For any inner vertices $p,q \in V_I^{in} \cup V_I^{*}$ and $u,v \in V_I^{out}$, the I-path from $p$ to $q$ should not intersect the I-path from $u$ to $v$.

        \item \textit{C2:} For any inner vertices $p,q \in V_I^{out} \cup V_I^{*}$ and $u,v \in V_I^{in}$, the I-path from $p$ to $q$ should not intersect the I-path from $u$ to $v$. \\
\end{itemize}

These  conditions are shown to be not necessary by an explicit example. From the code construction in \cite{ViR}, the length of the index code obtained for any IC structure with cycles among non-inner vertex set is $N-K+1$. Algorithm in \cite{SaR} gives an index code of length $N-K+2-t$, where $t$ is the maximum number of disjoint cycles in $\cal{C}$, for IC structures with interlocked outer cycles which obey either Condition $\textit{C1}$ or $\textit{C2}$. There is a difference of $t-1$ in the length of index codes. Hence $t$ is a critical parameter for IC structures with interlocked outer cycles, especially for the IC structures with $t$ more than one since it indicates the number of transmissions that can be saved whenever $t$ is greater than 1.

\subsection{Our Contributions}
In this paper we present a simple algorithm to find a set of maximum number of disjoint outer cycles in IC structures with interlocked outer cycles. The cardinality of this set gives $t$ which controls the optimal length of index code for IC structures with interlocked cycles.	
\section{Structure of IC Structures with Interlocked Outer Cycles}

	In this section, we describe some of the properties of IC structures with interlocked outer cycles. We consider  IC structures $G_K$ containing interlocked outer cycles with $N$ number of total vertices and $K$ number of inner vertices. Let \\
$V_I^{out}=\{k: k \in V_I$ and there exists a path from $k$ to $j$ for some $j \in V_{OC},$ where the path doesn't include any inner vertex other than $k\},$ \\
$V_I^{in}=\{k: k \in V_I$ and there exists a path from $j$ to $k$ for some $j \in V_{OC},$ where the path doesn't include any inner vertex other than $k\}$ and \\
$V_I^{*} = V_I \backslash (V_I^{in} \cup V_I^{out})$. \\
%
%
%
%
%
%

Next in a sequence of four lemmas we show that if all the $n$ cycles in $\cal{C}$ do not have a common vertex then the IC structure with interlocked cycles will have exactly one central cycle. This is an important property which leads to our algorithm to find the maximum number of disjoint cycles in $\cal{C}.$ The proofs for the first two lemmas, Lemma \ref{lem: unique_vertices_arcs_cyc} and Lemma \ref{lem: exist_in_and _out_paths_UoC}, are lengthy and given in appendix. \\

\begin{lem}
			Let $C_c$ be a central cycle in $C$. For any other cycle $C_k \in \cal{C}$, if $V_{C_k} \backslash V_{c,k} \neq \phi$, then there exist some vertices or arcs in $C_k$ which are unique to $C_k$. 
\label{lem: unique_vertices_arcs_cyc}
\end{lem}
~ \\

Table \ref{tab2} illustrates Lemma \ref{lem: unique_vertices_arcs_cyc} for all the five examples discussed in the previous section. \\

                        \begin{table*}
                                \begin{center}
                                        \begin{tabular}{ | p{1.5cm} | p{1.8cm} | p{2.5cm}  |p{3.5cm} |}
                                                \hline

                                                \textit{Example}& \textit{Central Cycle $C_c$}  &  \textit{Cycle $C_k$ other than $C_c$}  &  \textit{The path from $i \in V_{C_k}$ to $j \in V_{C_k}$ which is unique to $C_k$, except $i$ and $j$.}  \\ \hline

                                                \multirow{3}{*}{Example \ref{exmp12}}&
                                                \multirow{2}{*}{$C_1$} & $C_2$ & $(12 \rightarrow 11)$ \\ \cline{3-4} &
                                                                           & $C_3$ & $(12 \rightarrow 11)$ \\ \cline{2-4} &
                                                \multirow{2}{*}{$C_2$} & $C_1$ & $(12 \rightarrow 11)$ \\ \cline{3-4} &
                                                                       & $C_3$ & $(12 \rightarrow 11)$ \\ \cline{2-4} &
                                                \multirow{2}{*}{$C_3$} & $C_1$ & $(12 \rightarrow 11)$ \\ \cline{3-4} &
                                                                       & $C_2$ & $(12 \rightarrow 11)$ \\ \hline

                                                \multirow{3}{*}{Example \ref{exmp13}}&
                                                \multirow{2}{*}{$C_1$} & $C_2$ & Cycle $C_2$ except the vertex $8$ \\ \cline{3-4} &
                                                                   & $C_3$ & Cycle $C_3$ except the vertex $8$ \\ \cline{2-4} &
                                                \multirow{2}{*}{$C_2$} & $C_1$ & Cycle $C_1$ except the vertex $8$ \\ \cline{3-4} &
                                                                       & $C_3$ & Cycle $C_3$ except the vertex $8$ \\ \cline{2-4} &
                                                \multirow{2}{*}{$C_3$} & $C_1$ & Cycle $C_1$ except the vertex $8$ \\ \cline{3-4} &
                                                                       & $C_2$ & Cycle $C_2$ except the vertex $8$ \\ \hline

                                                \multirow{3}{*}{Example \ref{exmp14}}&
                                                \multirow{2}{*}{$C_1$} & $C_2$ & $(15 \rightarrow 13)$ \\ \cline{3-4} &
                                                                       & $C_3$ & $(15 \rightarrow 14)$ \\ \cline{2-4} &
                                                \multirow{2}{*}{$C_2$} & $C_1$ & $(14 \rightarrow 13)$ \\ \cline{3-4} &
                                                                       & $C_3$ & $(15 \rightarrow 14)$ \\ \cline{2-4} &
                                                \multirow{2}{*}{$C_3$} & $C_1$ & $(14 \rightarrow 13)$ \\ \cline{3-4} &
                                                                       & $C_2$ & $(15 \rightarrow 13)$ \\ \hline

                                                \multirow{3}{*}{Example \ref{exmp15}}&
                                                \multirow{3}{*}{$C_1$} & $C_2$ & $(13 \rightarrow 12)$ \\ \cline{3-4} &
                                                                       & $C_3$ & $(15 \rightarrow 14)$  \\ \cline{3-4} &
                                                                       & $C_4$ & Cycle $C_4$ except the vertex $13$ \\ \hline

                                                \multirow{3}{*}{Example \ref{exmp16}}&
                                                \multirow{3}{*}{$C_1$} & $C_2$ & $(22 \rightarrow 20)$ \\ \cline{3-4} &
                                                                       & $C_3$ & $(13 \rightarrow 22)$ \\ \cline{3-4} &
                                                                       & $C_4$ & $(19 \rightarrow 18)$ \\ \hline

                                        \end{tabular}

                                \end{center}
                                \caption{Table that illustrates Lemma \ref{lem: unique_vertices_arcs_cyc} for some Examples.}
                                                \label{tab2}
                        \end{table*}

		\begin{lem}
		Let $C_c$ be a central cycle in $C$. For any other cycle $C_k \in \cal{C}$,
			\begin{enumerate}

				\item	$\tilde{V}_{C_k} \neq \phi$,
				
				\item	 there exists atleast one path $P_{i,v_{k}}$ from some inner vertex $i \in V_I$ to some non-inner vertex $v_{k} \in \tilde{V}_{C_k}$, where the path doesn't include any vertex in $V_{OC}$ other than $v_k$ and 
				\item there exists atleast one path $P_{v_{k}^{'},j}$ from some non-inner vertex $v_{k}^{'} \in \tilde{V}_{C_k} $ to some inner vertex $j \in V_I$, where the path doesn't include any vertex in $V_{OC}$ other than $v_k^{'}$. \\ 
			\end{enumerate}
			
			\label{lem: exist_in_and _out_paths_UoC}
		\end{lem}
Table \ref{tab1} illustrates Lemma \ref{lem: exist_in_and _out_paths_UoC} for Examples \ref{exmp12}, \ref{exmp13}, \ref{exmp14}, \ref{exmp15} and \ref{exmp16}.	\\	


                \begin{table*}
                \begin{center}
                        \begin{tabular}{ | p{1.5cm} | p{1cm} | p{1cm}  |p{4cm} |p{4cm} |}
                                \hline

                                \textit{Example}& \textit{Central Cycle $C_c$}  &  \textit{Cycle $C_k$ other than $C_c$}  &  \textit{The path from $i \in V_I$ to $v_{k} \in \tilde{V}_{C_k}$, where the path doesn't include any vertex in $V_{OC}$ other than $v_k$, $(i \rightarrow v_k)$.} &  \textit{The path from $v'_{k} \in \tilde{V}_{C_k} $ to $j \in V_I$, where the path doesn't include any vertex in $V_{OC}$ other than $v'_{k}$, $(v'_k \rightarrow j)$. } \\ \hline

                                \multirow{3}{*}{Example \ref{exmp12}}&
                                \multirow{2}{*}{$C_1$} & $C_2$ & $(1 \rightarrow 10)$ & $(9 \rightarrow 4)$ \\ \cline{3-5} &
                                                                       & $C_3$ & $(2 \rightarrow 13)$ & $(14 \rightarrow 5)$ \\ \cline{2-5} &
                                \multirow{2}{*}{$C_2$} & $C_1$ & $(3 \rightarrow 15)$ & $(16 \rightarrow 6)$ \\ \cline{3-5} &
                                                                       & $C_3$ & $(2 \rightarrow 13)$ & $(14 \rightarrow 5)$ \\ \cline{2-5} &
                                \multirow{2}{*}{$C_3$} & $C_1$ & $(3 \rightarrow 15)$ & $(16 \rightarrow 6)$ \\ \cline{3-5} &
                                                                           & $C_2$ & $(1 \rightarrow 10)$ & $(9 \rightarrow 4)$ \\ \hline

                                \multirow{3}{*}{Example \ref{exmp13}}&
                                \multirow{2}{*}{$C_1$} & $C_2$ & $(2 \rightarrow 11)$ & $(12 \rightarrow 4)$ \\ \cline{3-5} &
                                                       & $C_3$ & $(1 \rightarrow 13)$ & $(14 \rightarrow 6)$ \\ \cline{2-5} &
                                \multirow{2}{*}{$C_2$} & $C_1$ & $(3 \rightarrow 9)$ & $(10 \rightarrow 5)$ \\ \cline{3-5} &
                                                   & $C_3$ & $(1 \rightarrow 13)$ & $(14 \rightarrow 6)$ \\ \cline{2-5} &
                                \multirow{2}{*}{$C_3$} & $C_1$ & $(3 \rightarrow 9)$ & $(10 \rightarrow 5)$ \\ \cline{3-5} &
                                                       & $C_2$ & $(2 \rightarrow 11)$ & $(12 \rightarrow 4)$ \\ \hline

                                \multirow{3}{*}{Example \ref{exmp14}}&
                                \multirow{2}{*}{$C_1$} & $C_2$ & $(1 \rightarrow 18)$ & $(19 \rightarrow 4)$ \\ \cline{3-5} &
                                                       & $C_3$ & $(2 \rightarrow 16)$ & $(17 \rightarrow 5)$ \\ \cline{2-5} &
                                \multirow{2}{*}{$C_2$} & $C_1$ & $(3 \rightarrow 11)$ & $(12 \rightarrow 6)$ \\ \cline{3-5} &
                                                   & $C_3$ & $(2 \rightarrow 16)$ & $(17 \rightarrow 5)$ \\ \cline{2-5} &
                                \multirow{2}{*}{$C_3$} & $C_1$ & $(3 \rightarrow 11)$ & $(12 \rightarrow 6)$ \\ \cline{3-5} &
                                                       & $C_2$ & $(1 \rightarrow 18)$ & $(19 \rightarrow 4)$ \\ \hline

                                \multirow{1}{*}{Example \ref{exmp15}}&
                                \multirow{3}{*}{$C_1$} & $C_2$ & $(1 \rightarrow 10)$ & $(11 \rightarrow 6)$ \\ \cline{3-5} &
                                                       & $C_3$ & $(3 \rightarrow 16)$ & $(17 \rightarrow 5)$ \\ \cline{3-5} &
                                                   & $C_4$ & $(2 \rightarrow 9)$ & $(18 \rightarrow 4)$ \\ \hline

                                \multirow{1}{*}{Example \ref{exmp16}}&
                                \multirow{3}{*}{$C_1$} & $C_2$ & $(3 \rightarrow 28)$ & $(29 \rightarrow 5)$ \\ \cline{3-5} &
                                                       & $C_3$ & $(1 \rightarrow 30)$ & $(31 \rightarrow 6)$ \\ \cline{3-5} &
                                                       & $C_4$ & $(2 \rightarrow 33)$ & $(32 \rightarrow 4)$ \\ \hline

                        \end{tabular}

                \end{center}
                \caption{Table that illustrates Lemma \ref{lem: exist_in_and _out_paths_UoC} for some Examples.}
                        \label{tab1}

                \end{table*}

{
{
{
{
		
		\begin{figure}[!t]
			\centering
			\includegraphics[width=20pc]{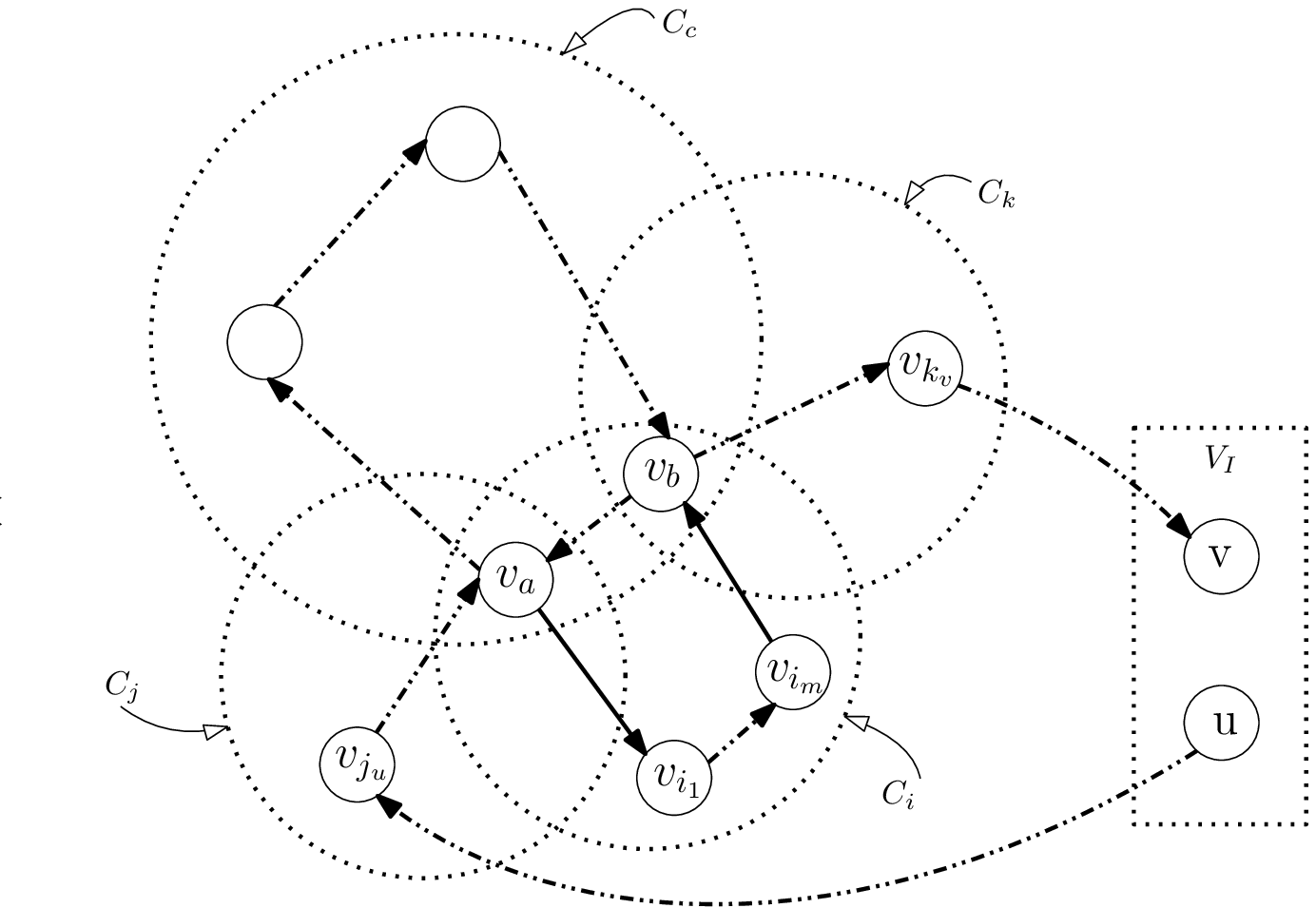}
			\caption{Illustration of the case in Lemma \ref{lem: com_vert_for_cycles}.}
			\label{fig: n_comm_vertex}
		\end{figure}

\begin{lem}
Let $C_c$ be a central cycle in $C$. For any other cycle $C_i \in \cal{C}$, all the cycles in $\cal{C}$ intersecting $C_i$, including $C_i$, have atleast a single vertex in common. If there is more than one central cycle, then all the $n$ cycles in the interlocked outer cycles $\cal{C}$ have atleast a single vertex in common. \label{lem: com_vert_for_cycles}
\end{lem}

\begin{IEEEproof}
On the contrary, let there exist $C_j, C_k \in \cal{C}$$ \backslash (C_c \cup C_i)$ such that $V_{C_j} \cap V_{C_i} \neq \phi, V_{C_k} \cap V_{C_i} \neq \phi$ and $V_{C_j} \cap V_{C_k} = \phi$, where $j\neq k \neq i$. By CCC, there exist some vertex $v_a \in V_{c,i} \cap V_{c,j}$ and some vertex $v_b \in V_{c,i} \cap V_{c,k}$. By ILC, if any two cycles intersect, then the intersection should have exactly a path in common. Hence, either the path from $v_a$ to $v_b$ or the path from $v_b$ to $v_a$ coincides in both $C_i$ and $C_c$ (since $v_a,v_b \in V_{c,i}$). Let the path from $v_b$ to $v_a$ coincides in both $C_i$ and $C_c$ (shown in Fig. \ref{fig: n_comm_vertex}). Then there exists a path $P_{v_a,v_b}$ from $v_a$ to $v_b$ in the cycle $C_i$ and a path $P_{v_a,v_b}^{'}$ from $v_a$ to $v_b$ in the cycle $C_c$ such that both the paths do not coincide. From Lemma \ref{lem: exist_in_and _out_paths_UoC}, there exists a path $P_{u,v_{j_u}}$ from some inner vertex $u \in V_I$ to some vertex $v_{j_u} \in \tilde{V}_{C_j}$ and a path $P_{v_{k_v},v}$ from some vertex $v_{k_v} \in \tilde{V}_{C_k}$ to some inner vertex $v \in V_I$. Let $P_{v_{j_u},v_a}$ be the path from $v_{j_u}$ to $v_a$ (part of the cycle $C_j$) and $P_{v_b,v_{k_v}}$ be the path from $v_b$ to $v_{k_v}$ (part of the cycle $C_k$). ($P_{u,v_{j_u}} \rightarrow P_{v_{j_u},v_a} \rightarrow P_{v_a,v_b} \rightarrow P_{v_b,v_{k_v}} \rightarrow P_{v_{k_v},v})$ forms an I-path from $u$ to $v$ and $(P_{u,v_{j_u}} \rightarrow P_{v_{j_u},v_a} \rightarrow P_{v_a,v_b}^{'} \rightarrow P_{v_b,v_{k_v}} \rightarrow P_{v_{k_v},v})$ forms another I-path from $u$ to $v$. Hence there exist two I-paths from $u$ to $v$. It contradicts the definition of an IC-structure. Hence for any cycle $C_i \in \cal{C}$$ \backslash C_c$, all the cycles in $\cal{C}$ intersecting $C_i$, including $C_i$, have atleast a single vertex in common.

                         If there is more than one central cycle, take one central cycle $C_c \in \cal{C}$. There exists some other central cycle $C_p \in \cal{C}$. $C_p, C_c$ and all other cycles in $\cal{C}$ have atleast a single vertex in common, since all the cycles in $\cal{C}$ intersect $C_p$. Hence if there is more than one central cycle, then all the $n$ cycles in the interlocked outer cycles $\cal{C}$ have atleast a single vertex in common.
\end{IEEEproof}

For the IC structure in Example \ref{exmp12} all the three outer cycles are central cycles and these three have the vertices $11$ and $12$ in common. For the IC structure in Example \ref{exmp13}  also all the three outer cycles are central cycles having the vertex $8$ in common. The same being the case for the IC structure of Example \ref{exmp14} with the vertex $14$ being common.   Consider Example \ref{exmp15}. For this IC structure, there is only one central cycle - $C_1$. Consider the cycle $C_2$. The outer cycles intersecting $C_2$ are $C_1$ and $C_4$. $C_2, C_1$ and $C_4$ have the vertex $13$ in common. The only cycle intersecting $C_3$ is $C_1$. $C_3$ and $C_1$ have the vertices $14$ and $15$ in common. Similar situation can be seen in Example \ref{exmp16}. For the IC structure of this example  there is only one central cycle - $C_1$. Consider the cycle $C_2$. The outer cycles intersecting $C_2$ are $C_1$ and $C_3$. $C_2, C_1$ and $C_3$ have the vertex $22$ in common. The only cycle intersecting $C_4$ is $C_1$. $C_4$ and $C_1$ have the vertices $18$ and $19$ in common. \\

\begin{lem}
If all the $n$ cycles in the interlocked outer cycles $\cal{C}$ do not have any vertex in common, then there exists exactly one central cycle in $\cal{C}$. \label{lem: unique_PCC}
\end{lem}

\begin{IEEEproof}
On the contrary, let there exist more than one central cycle. If there is more than one central cycle in $\cal{C}$, then take any central cycle $C_p$ in $\cal{C}$. For any other cycle $C_j \in \cal{C}$, the cycles $C_j,C_p$ and all the cycles in $\cal{C}$ intersecting $C_j$ have atleast a single vertex in common by Lemma \ref{lem: com_vert_for_cycles}. Since there is more than one central cycle, some other central cycle $C_q$ will be there in $\cal{C}$$\backslash C_p$. All other cycles in $\cal{C}$ intersect $C_q$. Hence $C_p,C_q$ and all other cycles in $\cal{C}$ have atleast a single vertex in common. That contradicts our case. Hence if all the $n$ cycles in the interlocked outer cycles $\cal{C}$ do not have any vertex in common, then there exists exactly one central cycle in $\cal{C}$. 
\end{IEEEproof}
	
For Examples \ref{exmp12}, \ref{exmp13} and \ref{exmp14},  all the three outer cycles are central cycles with common vertices being respectively $\{11,12\}$ $\{8\}$ and $\{14\}$.   

%
%

 For both the  Example \ref{exmp15} and Example \ref{exmp16} the  corresponding  IC structure has four outer cycles and they do not have any vertex in common. The only central cycle is $C_1$.
\begin{lem}
                        If all the $n$ cycles in the interlocked outer cycles $\cal{C}$ do not have any vertex in common, then let $C_c$ be the central cycle in $\cal{C}$. For any other cycle $C_j \in \cal{C}$, let $C_j^{*}$ be the set of cycles in $\cal{C}$$ \backslash \{C_c\}$ intersecting $C_j$. Each of the cycles in $\cal{C} $$\backslash (C_j^{*} \cup \{C_c\})$ is disjoint from each of the cycles in $C_j^{*}$.  \label{lem: dist_cyc_set}
\end{lem}

\begin{IEEEproof}
                        For each cycle $C_k \in C_j^{*}$, $C_j$ intersects $C_k$. By Lemma \ref{lem: com_vert_for_cycles}, all the cycles in $\cal{C}$ intersecting $C_k$, including $C_k$, have atleast a single vertex $v_b$ in common. Since $C_j$ is one of the cycles intersecting $C_k$, $C_j$ has the vertex $v_b$ in common with $C_k$ and all other cycles in $\cal{C}$$ \backslash \{C_c,C_k\}$ intersecting $C_k$. All the cycles in $\cal{C} $$\backslash \{C_c,C_j\}$ intersecting $C_j$ are included in $C_{j}^{*}$ and all the cycles in $\cal{C}$$ \backslash \{C_c,C_k\}$ intersecting $C_k$ intersect the cycle $C_j$. Hence all the cycles in $\cal{C}$$ \backslash \{C_c,C_k\}$ intersecting $C_k$ are included in $C_j^{*}\backslash C_k$. For each $C_k \in C_{j}^{*}$, none of the cycles in $\cal{C} $$\backslash \{C_j^{*} \cup C_c\}$ intersects $C_k$, since cycles intersecting $C_k$ intersect $C_j$ also. Hence each of the cycles in $\cal{C}$$ \backslash (C_j^{*} \cup \{C_c\})$ is disjoint from each of the cycles in $C_j^{*}$.
\end{IEEEproof}

                Consider Example \ref{exmp15}. For this IC structure, the only central cycle is $C_1$. The set $\{C_2,C_4\}$ and the set $\{C_3\}$ are disjoint.

                Consider Example \ref{exmp16}. For this IC structure, the only central cycle is $C_1$. The set $\{C_2,C_3\}$ and the set $\{C_4\}$ are disjoint.
		
                \section{Algorithm}

                In this section, we provide a simple algorithm to find a set of maximum number of disjoint outer cycles in IC structures with interlocked outer cycles.

Let $S$ be a set of minimum number of vertices to be removed from $V_{OC}$ to make all the cycles in $\cal{C}$ acyclic and $C^S$ is the set of maximum number of disjoint cycles among the non-inner vertex set. Also, let $|C^S|=t$. If all the $n$ cycles in $\cal{C}$ have atleast one vertex in common, then $t$ is one and $C^S= C_i$, for some $C_i \in \cal{C}$.

                If all the $n$ cycles in $\cal{C}$ do not have any vertex in common, then there exists exactly one central cycle in $\cal{C}$ by Lemma \ref{lem: unique_PCC}. Algorithm \ref{algo} finds $C^S$ if all the $n$ cycles in $\cal{C}$ do not have any vertex in common

                \begin{algorithm}
                        \caption{Algorithm to obtain a set of maximum number of disjoint outer cycles if all the outer cycles in an IC structure do not have any vertex in common.}
                        \SetKwInOut{Input}{Input}
                        \SetKwInOut{Output}{Output}
                        \Input{An IC structure $G_K$ with the central cycle $C_c$, the set $C^r=\cal{C}$$ \backslash C_c,V_{OC}$, a set $C^S$ and a set $S$, where $S$ is a set of minimum number of vertices to be removed from $V_{OC}$ to make all the cycles in $\cal{C}$ acyclic and $C^S$ is the set of maximum number of disjoint cycles among the non-inner vertex set. $S$ and $C^S$ are initialized as $\phi$.}
                        \Output{$C^S$ and $S$.}

                        \While{$C^{r} \neq \phi$}{
                        \begin{enumerate}
                        \item Choose any cycle $C_j \in C^{r}$. Find a vertex\\ $v_{j} \in V_{c,j}$ which is in common with $C_j$ and all \\the cycles intersecting $C_j$ (If there is more \\than one vertex, choose any one of them).

                        \item Remove the vertex $v_{j}$ from $V_{OC}$ and put it in \\the set $S$. Put $C_j$ in the set $C^{S}$. Remove all \\the cycles intersecting $v_{j}$ from $C^{r}$.
                \end{enumerate} \label{algo}
                        }

                \end{algorithm}

                \textit{Proof of Correctness of Algorithm \ref{algo}:}
                Consider an IC structure $G_K$. Let $C_c$ be the central cycle in $G_K$. Let $S$ be a set of minimum number of vertices to be removed from $V_{OC}$ to make all the cycles in $\cal{C}$ acyclic. Let $C^S$ be a set of maximum number of disjoint cycles in $\cal{C}$. Algorithm \ref{algo} finds $S$ and $C^S$ for $G_K$. Algorithm \ref{algo} is explained below.

                If the cardinality of the set $S$ is $t$, then the cardinality of the set $C^{S}$ is also $t$, since every time when we add a new vertex to $S$, we add a new cycle to $C^{S}$ also. Lemma \ref{lem: com_vert_for_cycles} guarantees that there exists a vertex $v_{j}$ chosen in step $1$, such that all the cycles intersecting $C_j$ passes through $v_{j}$. After choosing a cycle $C_j$ and a vertex $v_{j}$ in step $1$ we have removed the vertex $v_{j}$ in step $2$. Hence all the cycles intersecting $C_j$ and the cycle $C_j$ have become acyclic. Let $C_{j}^{*}$ be the set of all cycles in $\cal{C}$ passing through $v_{j}$, excluding $C_c$. From Lemma \ref{lem: dist_cyc_set}, none of the cycles (except $C_c$) in $\cal{C}$ outside the set $C_{j}^{*}$ intersect any of the cycles in $C_{j}^{*}$. Therefore, if we choose any cycle $C_k$ from the new $C^{r}$ obtained after removing the cycles passing through $v_{j}$, it doesn't intersect $C_j$. Hence all the cycles in the set $C^{S}$ are disjoint.

                $\hat{C}$ is the union of $C_c$ and $t$ sets of cycles, $C_{j}^{*}$. For each $C_j \in C^{S}$, $C_{j}^{*}$ represents the set of all cycles in $\cal{C}$ intersecting $C_j$ (including $C_j$ and excluding $C_c$). The $t$ sets of cycles, $C_{j}^{*}$, are disjoint from one another. All the cycles in the set $C_{j}^{*}$ have a common vertex $v_{j}$. So the maximum number of disjoint cycles we can get from a set is $1$ and the maximum number of disjoint cycles we can get from $t$ sets is $t$. Since all the cycles in $\cal{C}$ intersect $C_c$, the maximum number of disjoint cycles in $\cal{C}$ is $t$ and the minimum number of vertices to be removed from $V_{OC}$ to make the cycles in $\cal{C}$ acyclic is $t$. This completes the proof of correctness. \\
	
	For the IC structure of  Example \ref{exmp12}  $C^{S}=\{C_1\}, S=\{11\}$ and $t=1$ and for the IC structure of Example \ref{exmp13},  $C^{S}=\{C_1\}, S=\{8\}$ and $t=1$. Consider Example \ref{exmp14}. For the IC structure of this example $C^{S}=\{C_1\}, S=\{14\}$ and $t=1$. We have for the IC structure of  Example \ref{exmp15},  $C^{S}=\{C_2,C_3\}, S=\{13,14\}$ and $t=2$ and for the IC structure of 
 Example \ref{exmp16},  $C^{S}=\{C_3,C_4\}, S=\{22,19\}$ and $t=2$.

	\section{Conclusion}
		In this work, we have studied IC structures with interlocked outer cycles and have provided a simple  algorithm to find a set $C^S$ of maximum number of disjoint cycles in the non-inner vertex set. The cardinality of $C^S$ gives the maximum number of disjoint outer cycles $t$ which controls the optimal length of index code for IC structures with interlocked cycles when either the condition $C1$ or $C2$ is satisfied.
	
	
	\section*{Acknowledgment}
	This work was supported partly by the Science and Engineering Research Board (SERB) of Department of Science and Technology (DST), Government of India, through J. C. Bose National Fellowship to B. Sundar Rajan.
	


\begin{center}
APPENDIX A

Proof of Lemma 2.1

\end{center}
\label{append1}
                        Let $C_c$ be a central cycle in $C$. For any other cycle $C_k \in \cal{C}$, if $V_{C_k} \backslash V_{c,k} \neq \phi$, then on the contrary let there do not exist any vertices and arcs in $C_k$ which are unique to $C_k$ (shown in Fig. \ref{fig: n_arcs_vertices_not_null}).
                       
\begin{figure*}[!t]
                        \centering
                        \includegraphics[width=28pc]{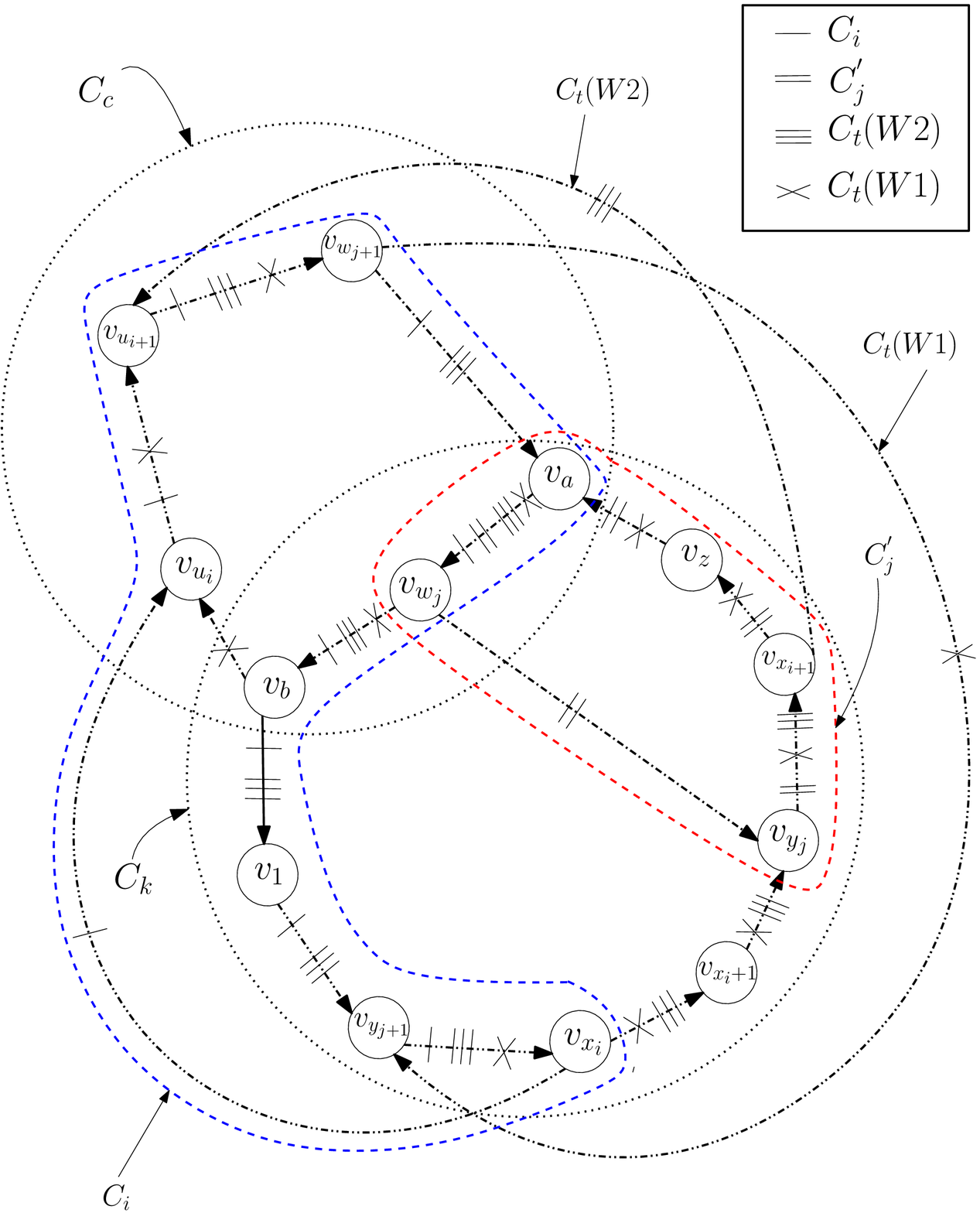}
                        \caption{Illustration of the case in Lemma \ref{lem: unique_vertices_arcs_cyc}.}
                        \label{fig: n_arcs_vertices_not_null}
                \end{figure*}

                \begin{figure}[!t]
                        \centering
                        \includegraphics[width=20pc]{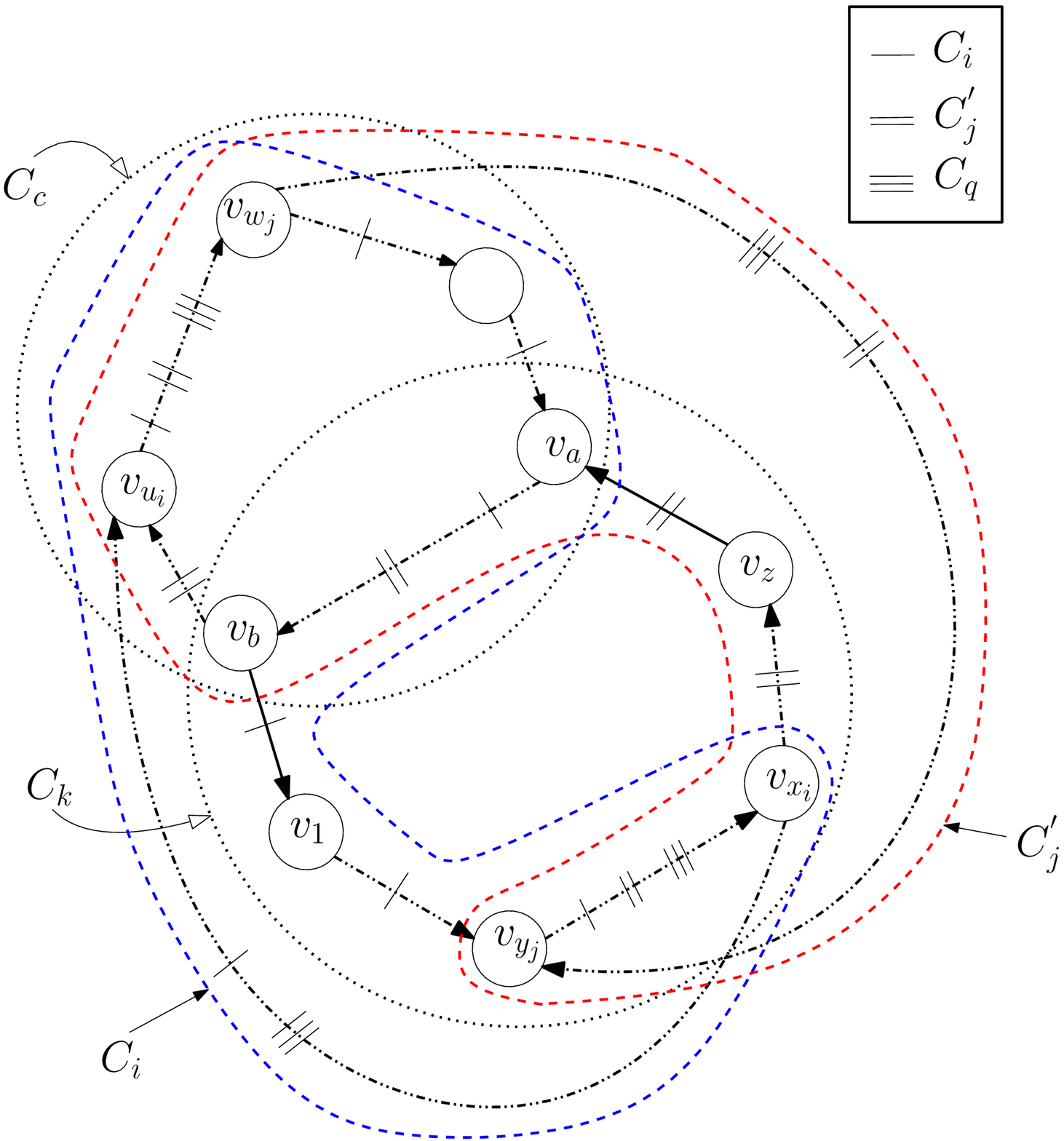}
                        \caption{Illustration of the case in Lemma \ref{lem: unique_vertices_arcs_cyc}.}
                        \label{fig: n_arcs_vertices_not_null_final}
                \end{figure}

                \begin{figure*}[!t]
                        \centering
                        \includegraphics[width=25pc]{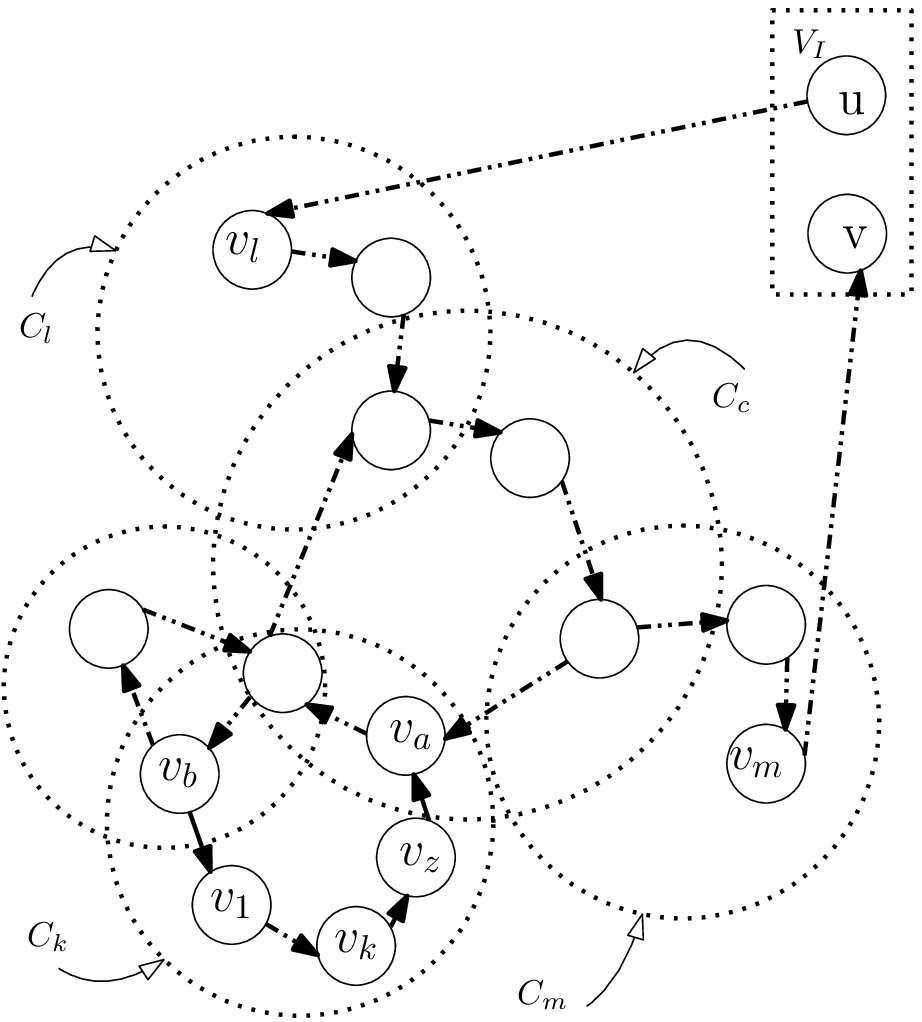}
                        \caption{Illustration of Claim 1 in Lemma \ref{lem: exist_in_and _out_paths_UoC}.}
                        \label{Fig: n_exist_path_1}
                \end{figure*}

                \begin{figure*}[!t]
                        \centering
                        \includegraphics[width=25pc]{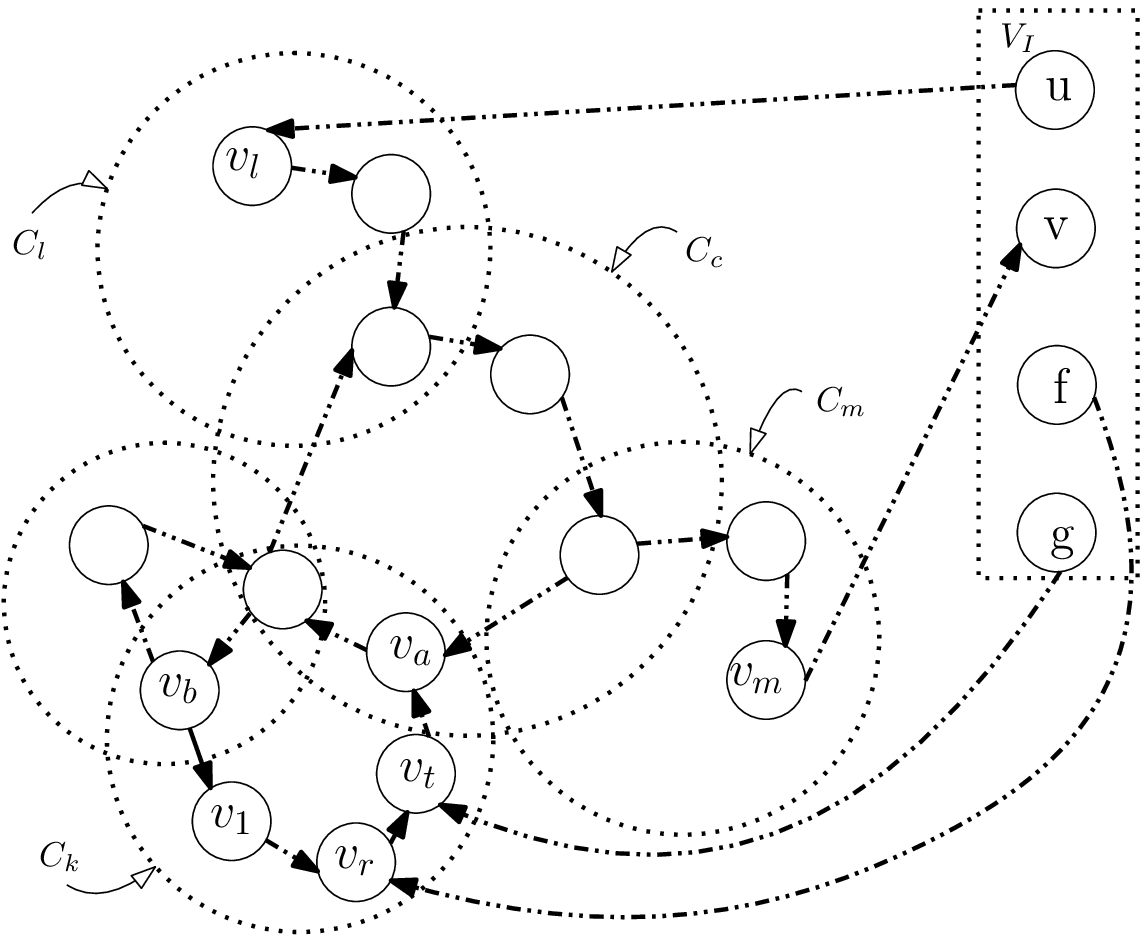}
                        \caption{Illustration of Claim 2 in Lemma \ref{lem: exist_in_and _out_paths_UoC}.}
                        \label{Fig: n_exist_path_2}
                \end{figure*}

                \begin{figure*}[!t]
                        \centering
                        \includegraphics[width=25pc]{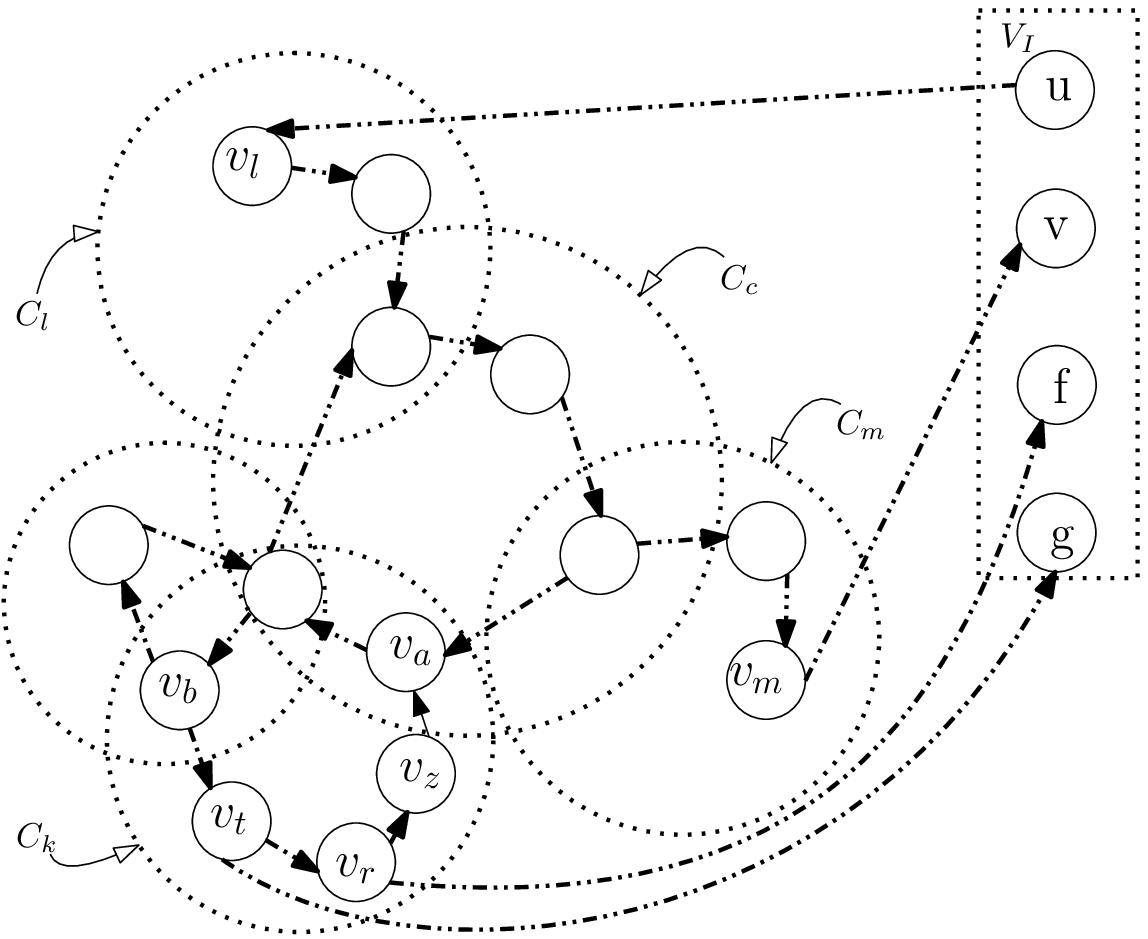}
                        \caption{Illustration of Claim 3 in Lemma \ref{lem: exist_in_and _out_paths_UoC}.}
                        \label{Fig: n_exist_path_3}
                \end{figure*}

                        If any two cycles intersect, from ILC, we know that they have exactly a path in common. Let the path $P_{v_a,v_b}$ from some vertex $v_a \in V_{c,k}$ to some other vertex $v_b \in V_{c,k}$ be common to both $C_k$ and $C_c$. $v_a$ and $v_b$ can coincide. Let $V_{C_k} \backslash V_{c,k} =\{v_1,v_2,...,v_z\}$ such that the path $P_{v_b,v_a}=(v_b \rightarrow v_1 \rightarrow v_2 \rightarrow.... \rightarrow v_z \rightarrow v_a)$ exists in $C_k$.

                        Let $i=1$ and $j=1$. From our assumption, the path $(v_b \rightarrow v_i)$ should be included in some cycle in $\cal{C}$. Let it be in $C_i$ ($C_i$ in Fig. \ref{fig: n_arcs_vertices_not_null}). From the definition of interlocked outer cycles, if any two cycles intersect, then the intersection should have exactly a path in common. Hence let $P_{v_{u_i},v_{b}}$ be the path from some vertex $v_{u_i} \in V_{C_c}$ to $v_b$ which is common to both $C_c$ and $C_i$. The vertices $v_{u_i}$ and $v_b$ can coincide. Let $\lceil r,t \rfloor$ be $t$ if $t \in V_{c,k}$, else let it be $r$, where $r,t \in V_{C_c}$. Let $P_{ \lceil v_a,v_{u_i} \rfloor, v_{x_i}}$ be the path from $\lceil v_a,v_{u_i} \rfloor$ to some vertex $v_{x_i} \in V_{C_k} \backslash V_{c,k}$ which is common to both $C_i$ and $C_k$. Hence there exists a path $P_{v_{x_i},v_{u_i}}$ in $C_i$ from $v_{x_i}$ to $v_{u_i}$ which doesn't include any other vertices in $C_k$ or $C_c$ other than $v_{x_i}$ and $v_{u_i}$. 
                        
                        Also, from our assumption, the path $(v_z \rightarrow v_a)$ should be included in some cycle in $\cal{C}$. Let it be in $C_j^{'}$ ($C_j^{'}$ in Fig. \ref{fig: n_arcs_vertices_not_null}). From the definition of interlocked outer cycles, if any two cycles intersect, then the intersection should have exactly a path in common. Hence let $P_{v_a,v_{w_j}}$ be the path from vertex $v_a$ to some vertex $v_{w_j} \in V_{C_c}$ which is common to both $C_c$ and $C_j^{'}$. The vertices $v_{w_j}$ and $v_a$ can coincide. Let $P_{ v_{y_j}, \lceil v_b, v_{w_j} \rfloor}$ be the path from some vertex $v_{y_j} \in V_{C_k} \backslash V_{c,k}$ to $\lceil v_b,v_{w_j} \rfloor$ which is common to both $C_j^{'}$ and $C_k$. Hence there exists a path $P_{v_{w_j},v_{y_j}}$ in $C_j^{'}$ from $v_{w_j}$ to $v_{y_j}$ which doesn't include any other vertices in $C_k$ or $C_c$ other than $v_{w_j}$ and $v_{y_j}$.
                        
                        If $v_{y_j}$ comes before $v_{x_j}$ in the path $P_{v_b,v_a}$, let $P_{v_{u_i},v_{w_j}}$ be the path from $v_{u_i}$ to $v_{w_j}$ (part of $C_c$) and $P_{v_{y_j},v_{x_i}}$ be the path from $v_{y_j}$ to $v_{x_i}$ (part of $C_k$). Since $v_{x_i}$ comes in the path from $v_{y_j}$ to $v_a$ in $C_k$, the path $P_{v_{y_j},v_{x_i}}$ doesn't intersect the cycle $C_c$. $(P_{v_{w_j},v_{y_j}} \rightarrow P_{v_{y_j},v_{x_i}} \rightarrow P_{v_{x_i},v_{u_i}} \rightarrow P_{v_{u_i},v_{w_j}})$ forms some cycle $C_q$. The path $P_{v_{y_j},v_{x_i}}$ is common to both $C_q$ and $C_k$. None of the vertices in $P_{v_{y_j},v_{x_i}}$ are included in $C_c$. Hence CCC is violated and interlocked outer cycles do not contain $C_q$.
                        
                        If the path from $v_{u_{i}}$ to $v_{w_{j}}$ in $C_c$ and the path $P_{v_a,v_b}$ do not intersect, let $P_{v_{u_i},v_{w_j}}$ be the path from $v_{u_i}$ to $v_{w_j}$ (part of $C_c$) and $P_{v_{y_j},v_{x_i}}$ be the path from $v_{y_j}$ to $v_{x_i}$ (part of $C_k$). $(P_{v_{w_j},v_{y_j}} \rightarrow P_{v_{y_j},v_{x_i}} \rightarrow P_{v_{x_i},v_{u_i}} \rightarrow P_{v_{u_i},v_{w_j}})$ forms some cycle $C_p$. The path $P_{v_{u_i},v_{w_j}}$ and the path $P_{v_a,v_b}$ are common to both $C_p$ and $C_k$. Since $P_{v_{u_i},v_{w_j}}$ and $P_{v_a,v_b}$ are disjoint, ILC is violated and interlocked outer cycles do not contain $C_p$.
                        
                        If $v_{y_j}$ comes after $v_{x_j}$ in the path $P_{v_b,v_a}$ and if the path from $v_{u_{i}}$ to $v_{w_{j}}$ in $C_c$ and the path $P_{v_a,v_b}$ intersect, then do the following.

                        \begin{itemize}
                                \item \textit{Step 1:} From our assumption, the path $(v_{x_i} \rightarrow v_{x_i+1})$ should be included in some cycle in $\cal{C}$. Let it be in $C_{t}$. $C_{t}$ can intersect $C_k$ and $C_c$ in two different ways (either $W1$ or $W2$) such that the path $(v_{x_i} \rightarrow v_{x_i+1})$ is covered in $C_{t}$.
                                \begin{itemize}
                                        
                                        \item \textit{W1:} The path $P_{v_a,v_{w_{j+1}}}$ from $v_a$ to some vertex $v_{w_{j+1}} \in V_{C_c}$ is common to both $C_c$ and $C_{t}$ ($C_t(W1)$ in Fig. \ref{fig: n_arcs_vertices_not_null}). The vertices $v_{w_{j+1}}$ and $v_a$ can coincide. The path $P_{ v_{y_{j+1}}, \lceil v_b, v_{w_{j+1}} \rfloor}$ from some vertex $v_{y_{j+1}} \in V_{C_k} \backslash V_{c,k}$ to $\lceil v_b,v_{w_{j+1}} \rfloor$ is common to both $C_{t}$ and $C_k$.
                                        \item \textit{W2:} The path $P_{v_{u_{i+1}},v_{b}}$ from some vertex $v_{u_{i+1}} \in V_{C_c}$ to $v_b$ is common to both $C_c$ and $C_{t}$ ($C_t(W2)$ in Fig. \ref{fig: n_arcs_vertices_not_null}). The vertices $v_{u_{i+1}}$ and $v_b$ can coincide. The path $P_{\lceil v_a,v_{u_{i+1}} \rfloor, v_{x_{i+1}}}$ from $\lceil v_a,v_{u_{i+1}} \rfloor$ to some vertex $v_{x_{i+1}} \in V_{C_k} \backslash V_{c,k}$ is common to both $C_{t}$ and $C_k$.
                                \end{itemize}
                                \item \textit{Step 2:} If $C_{t}$ intersects $C_k$ and $C_c$ as in $W1$, then let $j=j+1$ and exit from the steps.
                                \item \textit{Step 3:} If $C_{t}$ intersects $C_k$ and $C_c$ as in $W2$ and if $v_{x_{i+1}}$ comes in the path from $v_{y_j}$ to $v_a$ in $C_k$, then let $i=i+1$ and exit from the steps.
                                \item \textit{Step 4:} If $C_{t}$ intersects $C_k$ and $C_c$ as in $W2$ and if the path from $v_{u_{i+1}}$ to $v_{w_{j}}$ in $C_c$ and the path $P_{v_a,v_b}$ are disjoint, then let $i=i+1$ and exit from the steps.
                                \item \textit{Step 5:} Let $i=i+1$. Rename $C_t$ as $C_{i}$ and goto Step $1$.  
                        \end{itemize}

                        If $j=2,$ then rename $C_t$ as $C_{j}^{'}$, else rename $C_t$ as $C_{i}$. There exists a path $P_{v_{x_i},v_{u_i}}$ in $C_{i}$ from $v_{x_i}$ to $v_{u_i}$ which doesn't include any other vertices in $C_k$ or $C_c$ other than $v_{x_i}$ and $v_{u_i}$. Also, there exists a path $P_{v_{w_j},v_{y_j}}$ in $C_j^{'}$ from $v_{w_j}$ to $v_{y_j}$ which doesn't include any other vertices in $C_k$ or $C_c$ other than $v_{w_j}$ and $v_{y_j}$ (shown in Fig. \ref{fig: n_arcs_vertices_not_null_final}).

                        Since $v_{u_i},v_{w_j} \in V_{C_c}$, there exists a path $P_{v_{u_i},v_{w_j}}$ from $v_{u_i}$ to $v_{w_j}$ (part of $C_c$). Since $v_{y_j},v_{x_i} \in V_{C_k}$, there exists a path $P_{v_{y_j},v_{x_i}}$ from $v_{y_j}$ to $v_{x_i}$ (part of $C_k$). If we exit in either Step $2$ or $3$, $v_{x_i}$ comes in the path from $v_{y_j}$ to $v_a$ in $C_k$. Hence the path $P_{v_{y_j},v_{x_i}}$ doesn't intersect the cycle $C_c$. The vertices $v_{x_i}$ and $v_{y_j}$ can coincide. $(P_{v_{w_j},v_{y_j}} \rightarrow P_{v_{y_j},v_{x_i}} \rightarrow P_{v_{x_i},v_{u_i}} \rightarrow P_{v_{u_i},v_{w_j}})$ forms some cycle $C_q$. The path $P_{v_{y_j},v_{x_i}}$ is common to both $C_q$ and $C_k$. None of the vertices in $P_{v_{y_j},v_{x_i}}$ are included in $C_c$. Hence CCC is violated and interlocked outer cycles do not contain $C_q$. Else if we exit in step $4$, $(P_{v_{w_j},v_{y_j}} \rightarrow P_{v_{y_j},v_{x_i}} \rightarrow P_{v_{x_i},v_{u_i}} \rightarrow P_{v_{u_i},v_{w_j}})$ forms some cycle $C_p$. The path $P_{v_{u_i},v_{w_j}}$ and the path $P_{v_a,v_b}$ are common to both $C_p$ and $C_k$. Since $P_{v_{u_i},v_{w_j}}$ and $P_{v_a,v_b}$ are disjoint, ILC is violated and interlocked outer cycles do not contain $C_p$.
                        
                        Hence for any other cycle $C_k \in \cal{C}$, if $V_{C_k} \backslash V_{c,k} \neq \phi$, then there exist some vertices or arcs in $C_k$ which are unique to $C_k$.
 \\


\pagebreak

\begin{center}
APPENDIX B

Proof of Lemma 2.2
\end{center}
\label{append2}                                        
                        \begin{claim}
                                I-path from $u$ to $v$, for any $u,v \in V_I$, doesn't include any of the vertices and arcs in $C_k$ which are unique to $C_k$ for the below two cases.
                                \begin{itemize}
                                        \item \textit{Case 1:} For $\tilde{V}_{C_k} = \phi$.
                                        
                                        \item \textit{Case 2:} For $\tilde{V}_{C_k} \neq \phi$, there doesn't exist any path $P_{i,v_{k}}$ from any $i \in V_I$ to any vertex $v_{k} \in \tilde{V}_{C_k}$, where the path doesn't include any vertex in $V_{OC}$ other than $v_k$ and there doesn't exist any path $P_{v'_{k},j}$ from any vertex $v'_{k} \in \tilde{V}_{C_k}$ to any $j \in V_I$, where the path doesn't include any vertex in $V_{OC}$ other than $v'_k$.
                                \end{itemize} \label{claim: 1}
                        \end{claim}
                        \begin{IEEEproof}
                                Let $C_c$ be a central cycle in $C$. From Lemma \ref{lem: unique_vertices_arcs_cyc}, if $V_{C_k} \backslash V_{c,k} \neq \phi$ for any other cycle $C_k \in \cal{C}$, then there exist some vertices or arcs in $C_k$ which are unique to $C_k$. Hence $\tilde{V}_{C_k} = \phi$ implies there is only a single arc in $C_k$ which is unique to $C_k$.   
                                
                                For both the cases, since there exists an I-path from some vertex $u \in V_I$ to some other vertex $v \in V_I$ which intersect some of the cycles in $\cal{C}$, let $P_{u,v_{l}}$ be the path from $u$ to some vertex $v_{l} \in V_{OC} \backslash \tilde{V}_{C_k}$, where the path doesn't include any vertex in $V_{OC}$ other than $v_l$ and $P_{v_{m},v}$ be the path from some vertex $v_{m} \in V_{OC} \backslash \tilde{V}_{C_k}$ to $v$, where the path doesn't include any vertex in $V_{OC}$ other than $v_m$ (shown in Fig. \ref{Fig: n_exist_path_1}). Since there exists a path from any vertex in $V_{OC}$ to any other vertex in $V_{OC}$ by $\textit{P2}$, let $P_{v_{l},v_{m}}$ be the path from the vertex $v_{l}$ to $v_{m}$. Since $v_{l},v_{m} \notin \tilde{V}_{C_k}$, the vertex $v_l$ will be included in some cycle $C_l \in \cal{C}$$ \backslash C_k$ and the vertex $v_m$ will be included in some cycle $C_m \in \cal{C}$$ \backslash C_k$. From $\textit{P2}$, the path from any vertex $v_{l} \in V_{C_l}$ to any other vertex $v_{m} \in V_{C_m}$ includes the vertices and arcs from $C_l\cup C_m\cup C_c$. Hence $(P_{u,v_{l}} \rightarrow P_{v_{l},v_{m}} \rightarrow P_{v_{m},v})$ forms an I-path from $u$ to $v$ and the I-path doesn't contain any of the vertices and arcs in $C_k$ which are not included in any other cycle in $\cal{C}$. Hence I-path between any two vertices $u,v \in V_I$ doesn't contain any of the vertices and arcs in $C_k$ which are unique to $C_k$.
                        \end{IEEEproof} 
                        
                        By Claim \ref{claim: 1}, some of the arcs and vertices will not be included in any of the I-paths, if $\tilde{V}_{C_k} = \phi$. Hence they won't be included in the union of rooted trees. That contradicts the definition of an IC structure. Hence $\tilde{V}_{C_k} \neq \phi$. For any cycle $C_k \in \cal{C} \backslash$$ C_c$ there exist $v_a,v_b \in V_{C_k} \backslash \tilde{V}_{C_k}$ and $\tilde{V}_{C_k}=\{v_1,v_2,...,v_z\}$ (since $\tilde{V}_{C_k} \neq \phi$), such that the path $P_{v_b,v_a}=(v_b \rightarrow v_1 \rightarrow v_2 \rightarrow...\rightarrow v_z \rightarrow v_a)$ exists in the cycle $C_k$. 

                        \begin{claim}
                                For $\tilde{V}_{C_k} \neq \phi$, let $v_{r} \in \tilde{V}_{C_k}$ such that some path $P_{f,v_{r}}$ from some $f \in V_I$ to $v_{r}$ exists which doesn't include any vertex in $V_{OC}$ other than $v_r$ and there does not exist any path from any of the inner vertices to any of the vertices in the path $P_{v_b,v_r}$ from $v_b$ to $v_r$ (except $v_b$ and $v_r$). If there exists a path $P_{i,v_{k}}$ from some $i \in V_I$ to some vertex $v_{k} \in \tilde{V}_{C_k}$, where the path doesn't include any vertex in $V_{OC}$ other than $v_k$ and there doesn't exist any path $P_{v'_{k},j}$ from any $v'_{k} \in \tilde{V}_{C_k}$ to any $j \in V_I$, where the path doesn't include any vertex in $V_{OC}$ other than $v'_{k}$, then none of the vertices and arcs in the path $P_{v_b,v_{r}}$ (except $v_b$ and $v_{r}$) will be included in any of the I-paths.

                        \end{claim}
                        
                        \begin{IEEEproof}       
                                For any I-path from some $u \in V_I$ to $v \in V_I$, where the path $P_{u,v_{l}}$ from $u$ to some vertex $v_{l} \in V_{OC} \backslash \tilde{V}_{C_k}$ which doesn't include any vertex in $V_{OC}$ other than $v_l$ and the path $P_{v_{m},v}$ from some vertex $v_{m} \in V_{OC} \backslash \tilde{V}_{C_k}$ to $v$ which doesn't include any vertex in $V_{OC}$ other than $v_m$, exist, we have shown in the proof of Claim \ref{claim: 1} that the I-path doesn't include any of the vertices and arcs in $C_k$ which are unique to $C_k$.
                                
                                For any I-path from some $g \in V_I$ to $v \in V_I$, where the path $P_{g,v_{t}}$ from $g$ to some vertex $v_{t} \in \tilde{V}_{C_k}$ which doesn't include any vertex in $V_{OC}$ other than $v_t$ and the path $P_{v_{m},v}$ from some vertex $v_{m} \in V_{OC} \backslash \tilde{V}_{C_k}$ to $v$ which doesn't include any vertex in $V_{OC}$ other than $v_m$, exist, we will show that none of the vertices and arcs in the path $P_{v_b,v_{r}}$ (except $v_b$ and $v_{r}$) will be included in the I-path (shown in Fig \ref{Fig: n_exist_path_2}).
                                
                                The path $P_{v_t,v_a}$ from $v_t$ to $v_a$ doesn't include any of the vertices and arcs in the path $P_{v_b,v_{r}}$ (except $v_b$ and $v_{r}$), since $v_t$ comes after the vertex $v_r$ in the path $P_{v_b,v_a}$. 
                                
                                Since there exists a path from any vertex in $V_{OC}$ to any other vertex in $V_{OC}$ by $\textit{P2}$, let $P_{v_{a},v_{m}}$ be the path from the vertex $v_{a}$ to $v_{m}$. Since $v_{m} \notin \tilde{V}_{C_k}$, the vertex $v_m$ will be included in some cycle $C_m \in \cal{C}$$ \backslash C_k$. From $\textit{P2}$, the path from $v_{a}$ to $v_{m} \in V_{C_m}$ includes the vertices and arcs from $ C_m\cup C_c$. Hence the path $P_{v_a,v_m}$ doesn't include any of the vertices and arcs in the path $P_{v_b,v_{r}}$ (except $v_b$ and $v_r$). Hence $(P_{g,v_{t}} \rightarrow P_{v_{t},v_{a}} \rightarrow P_{v_a,v_m} \rightarrow P_{v_{m},v})$ forms an I-path from $g$ to $v$ and the I-path doesn't contain any of the vertices and arcs in the path $P_{v_b,v_{r}}$ (except $v_b$ and $v_{r}$).
                                
                                Hence I-path between any two vertices $u,v \in V_I$ doesn't contain any of the vertices and arcs in the path $P_{v_b,v_{r}}$ (except $v_b$ and $v_{r}$).

                        \end{IEEEproof}

                        \begin{claim}
                                For $\tilde{V}_{C_k} \neq \phi$, let $v_{r} \in \tilde{V}_{C_k}$ such that some path $P_{v_{r},f}$ from $v_{r}$ to some $f \in V_I$ exists which doesn't include any vertex in $V_{OC}$ other than $v_r$ and there does not exist any path from any of the vertices in the path $P_{v_r,v_a}$ from $v_r$ to $v_a$ (except $v_a$ and $v_r$) to any of the inner vertices. If there exists a path $P_{v'_{k},j}$ from some vertex $v'_{k} \in \tilde{V}_{C_k}$ to some $j \in V_I$, where the path doesn't include any vertex in $V_{OC}$ other than $v'_{k}$ and there doesn't exist any path $P_{i,v_{k}}$ from any $i \in V_I$ to  any $v_{k} \in \tilde{V}_{C_k}$, where the path doesn't include any vertex in $V_{OC}$ other than $v_k$, then none of the vertices and arcs in the path $P_{v_r,v_{a}}$ (except $v_a$ and $v_{r}$) will be included in any of the I-paths.

                                \label{claim: 3}
                        \end{claim}
                        \begin{IEEEproof}
                                For any I-path from some $u \in V_I$ to $v \in V_I$, where the path $P_{u,v_{l}}$ from $u$ to some vertex $v_{l} \in V_{OC} \backslash \tilde{V}_{C_k}$ which doesn't include any vertex in $V_{OC}$ other than $v_l$ and the path $P_{v_{m},v}$ from some vertex $v_{m} \in V_{OC} \backslash \tilde{V}_{C_k}$ to $v$ which doesn't include any vertex in $V_{OC}$ other than $v_m$, exist, we have shown in Claim \ref{claim: 1} that the I-path doesn't include any of the vertices and arcs in $C_k$ which are unique to $C_k$.
                                
                                For any I-path from some $u \in V_I$ to $g \in V_I$, where the path $P_{u,v_{l}}$ from $u$ to some vertex $v_{l} \in V_{OC} \backslash \tilde{V}_{C_k}$ which doesn't include any vertex in $V_{OC}$ other than $v_l$ and the path $P_{v_{t},g}$ from some vertex $v_{t} \in \tilde{V}_{C_k}$ to $g$ which doesn't include any vertex in $V_{OC}$ other than $v_t$, exist, we will show that none of the vertices and arcs in the path $P_{v_r,v_{a}}$ (except $v_a$ and $v_{r}$) will be included in the I-path (shown in Fig \ref{Fig: n_exist_path_3}).
                                
                                The path $P_{v_b,v_t}$ from $v_b$ to $v_t$ doesn't include any of the vertices and arcs in the path $P_{v_r,v_{a}}$ (except $v_a$ and $v_{r}$), since $v_t$ comes before the vertex $v_r$ in the path $P_{v_b,v_a}$.
                                
                                Since there exists a path from any vertex in $V_{OC}$ to any other vertex in $V_{OC}$ by $\textit{P2}$, let $P_{v_{l},v_{b}}$ be the path from the vertex $v_{l}$ to $v_{b}$. Since $v_{l} \notin \tilde{V}_{C_k}$, the vertex $v_l$ will be included in some cycle $C_l \in \cal{C}$$ \backslash C_k$. From $\textit{P2}$, the path from $v_{l}$ to $v_{b}$ includes the vertices and arcs from $ C_l\cup C_c$. Hence the $P_{v_l,v_b}$ doesn't include any of the vertices and arcs in the path $P_{v_r,v_{a}}$ (except $v_a$). Hence $(P_{u,v_{l}} \rightarrow P_{v_{l},v_{b}} \rightarrow P_{v_b,v_t} \rightarrow P_{v_{t},g})$ forms an I-path from $u$ to $g$ and the I-path doesn't contain any of the vertices and arcs in the path $P_{v_r,v_{a}}$ (except $v_a$ and $v_{r}$). 
                                
                                Hence I-path between any two vertices $u,v \in V_I$ doesn't contain any of the vertices and arcs in the path $P_{v_r,v_{a}}$ (except $v_a$ and $v_{r}$).

                        \end{IEEEproof}                 
                        \textit{Proof (of Lemma \ref{lem: exist_in_and _out_paths_UoC}):} In all of the above claims atleast some of the arcs or vertices will not be included in any of the I-paths. Hence they won't be included in the union of rooted trees. That contradicts the definition of an IC structure. Hence $\tilde{V}_{C_k} \neq \phi$ and there exists atleast one path $P_{i,v_{k}}$, where the path doesn't include any vertex in $V_{OC}$ other than $v_k$ and atleast one path $P_{v'_{k},j}$, where the path doesn't include any vertex in $V_{OC}$ other than $v'_{k}$, for each $C_k \in \cal{C}$$ \backslash C_c$. 
                


\end{document}